\documentclass{sf2a-conf2022}
\usepackage{graphicx}
\usepackage{hyperref}
\usepackage[]{natbib}
\usepackage{mathrsfs}
\usepackage{amsmath}
\usepackage{amssymb}
\usepackage{epstopdf}

\def\BibTeX{{\rm B\kern-.05em{\sc i\kern-.025em b}\kern-.08em
    T\kern-.1667em\lower.7ex\hbox{E}\kern-.125emX}}
\bibpunct{(}{)}{;}{a}{}{,}  %%%%%%%%%%%%%  A&A bibliography style
\begin{document}

\TitreGlobal{SF2A 2022}

%%-----------------------------------------------------------------
%%      the top matter
%%

\title{Impact of magnetism on gravitational waves emitted\\by compact galactic binaries in quasi-circular orbits}

\runningtitle{Magnetism and compact galactic binaries in quasi-circular orbits}

\author{A. Bourgoin$^{1,}$}\address{SYRTE, Observatoire de Paris, PSL Research University, CNRS, Sorbonne Universit\'e, UPMC Univ. Paris 06, LNE, 61 avenue de l'Observatoire, 75014 Paris, France}
\address{Département d'Astrophysique-AIM, CEA, CNRS, Université Paris-Saclay, Université Paris Cité, 91191 Gif-sur-Yvette, France}

\author{E. Savalle}\address{Département de Physique des Particules, CEA, CNRS, Université Paris-Saclay, Université Paris Cité, 91191 Gif-sur-Yvette, France}

\author{C. Le Poncin-Lafitte$^1$}

\author{S. Mathis$^2$}

\author{M.-C. Angonin$^1$}

\author{A. Strugarek$^2$}

%% IF Author3 has the same affiliation than Author1:
%\author{C.\,E. Author3$^1$}

%% IF Author3 has its own affiliation:
%\author{C.\,E. Author3}\address{Dept. of Chess, University of Games, 35101 Las Vegas, Monaco} 

%% IF Author3 has two affiliations, the one of Author1 and a second one:
%\author{C.\,E. Author3$^{1,}$}\address{Dept. of Chess, University of Games, 35101 Las Vegas, Monaco} 

%% Keep this line, even if the page will be settled afterwards.
\setcounter{page}{237}

%%-----------------------------------------------------------------

\maketitle

%%-----------------------------------------------------------------
%%        The abstract
%% 
%%  Warning!  within the abstract:
%%  - do not use macros. 
%%  - do not use commands like: \cite, \citet, \citep ... etc.

\begin{abstract}
  The LISA (Laser Interferometer Space Antenna) mission will observe in the low frequency band from $0.1\,\mathrm{mHz}$ to $1\,\mathrm{Hz}$.  In this regime, we expect the galactic binaries to be the dominant (by number) sources of gravitational waves signal. Considering that galactic binaries are composed of the most magnetized astrophysical objects in the universe (i.e., the white dwarfs and the neutron stars), LISA is expected to bring new informations about the origin and the nature of magnetism inside degenerated stars. Currently, the data processing assumes that the galactic binary systems are non-magnetic and in circular orbits which can potentially biased the determination of the parameters of the sources and also the calibration of the detector. In this work, we investigate the impact of magnetism on gravitational waves emitted by compact galactic binaries assuming quasi-circular orbits.
\end{abstract}

%% Insert the keywords (to appear in the ADS indexing)
%% Keywords must be separated by a comma

%%-----------------------------------------------------------------

\section{Introduction}
%%---------------------

By observing the low frequency bandwidth of the gravitational waves (GWs) spectrum (i.e., from $0.1\,\mathrm{mHz}$ to above $1\,\mathrm{Hz}$), the Laser Interferometer Space Antenna (LISA) mission will bring precious informations concerning the galactic binaries (GBs) \citep{2017arXiv170200786A}. Indeed, it is expected that the mission will simultaneously resolve more than ten thousand of these binary systems \citep{PhysRevD.73.122001}. Among these, tenth of them, called ``\emph{verification binaries}'', are already known from electromagnetic observations and are identified as bright sources of GWs for LISA. They will serve for calibrating the sensitivity of the detector at first. Therefore, an error in modelling the signal from the verification binaries can potentially skew the determination of other extra-galactic sources observed by LISA. Currently, the data processing of the verification binaries assumes that GBs are quasi-monochromatic sources of GWs, which corresponds to the assumption that binary systems are inspiraling on circular orbits \citep{LDCGroup}. However, GBs comprise white dwarfs (WDs) and neutron stars (NSs), which both exhibit complex internal processes and intense magnetic fields (up to $10^9\,\mathrm{G}$ for WDs and up to $10^{15}\,\mathrm{G}$ for NSs; see e.g., \citet{2020AdSpR..66.1025F}). Internal physics and magnetism can thus change the quasi-monochromatic picture of the GWs signal detected by LISA and hence biased the calibration of the detector or the determination of the physical parameters (e.g., masses, semi-major axis, etc.). As a matter of fact, it was recently shown by \citet{2022PhRvD.105l4042B} that magnetism shifts all the frequencies present in the GWs signal. Therefore, in principle, LISA could let to measure the magnetism within thousand of binary systems and let to learn more about the origin and the nature of magnetism in degenerate stars by complementing spectropolarimetric electromagnetic observations \citep{2022ApJ...935L..12B}.

After recalling previous results from \citet{2022PhRvD.105l4042B} concerning the impact of magnetism on GWs emitted by a binary system of degenerate stars in quasi-circular orbits, we comment on how magnetism and eccentricity would be processed while making use of the quasi-monochromatic algorithms of the LISA datacode challenge (LDC) \citep{LDCGroup}.

\section{Orbital motion of a binary system and GWs emission}
%%-------------------------

In \citet{2022PhRvD.105l4042B}, the authors considered two compact and well-separated degenerate stars in a binary system of total mass $m=m_1+m_2$ with $m_1$ and $m_2$ the masses of the primary and secondary, respectively. The secular evolution of the system is derived assuming a relativistic description of the point-mass dynamics up to the 2.5 post-Newtonian (PN) approximation \citep{2014LRR....17....2B} which is coherent with the fact that orbital energy is radiated away from the source by GWs. Magnetic effects are considered within the magnetostatic approximation through the dipole-dipole magnetic interaction which is coherent with the ``\emph{fossil fields}'' hypothesis \citep{2005MNRAS.356..615F}. In addition, following \citet{2015MNRAS.454L...1S}, it was assumed that both the direction of the magnetic dipole moments of the primary and secondary (labeled $\mu_1$ and $\mu_2$) are aligned with the direction of their spins. Within this framework, it was shown by solving simultaneously the orbital and rotational motions, that the magnetic dipole-dipole interaction generates secular drifts on the mean longitude $(L)$ and on the longitude of the pericenter $(\varpi)$ which are given by \citep{2022PhRvD.105l4042B}
\begin{align}
  \dot\varpi_{\mathrm{M}}&=\frac{3\mu_0}{4\pi\sqrt{G}}\,\frac{\sqrt{m}}{m_1m_2}\,\frac{\mu_1\mu_2}{a^{7/2}}\,\cos\epsilon_1\cos\epsilon_2+\mathcal{O}(e^2)\text{,}\\
  \dot{L}_{\mathrm{M}}&=2\dot\varpi_{\mathrm{M}}+\mathcal{O}(e^2)\text{,}
\end{align}
where $G$ is the gravitational constant, $\mu_0$ is the magnetic permittivity of vacuum, $a$ is the semi-major axis of the orbit, $e$ is the eccentricity of the orbit, and $\epsilon_1$ and $\epsilon_2$ are the obliquities of the magnetic dipole moments with respect to the orbital plane.

From the secular evolution of the orbital motion, it is then possible to determine the mode polarizations of the GWs signal by making use of Einstein's well-know quadrupole formula. The expression for the GWs mode polarization $h_+$ (a similar expression for the GWs mode $h_\times$ exists) is given by
\begin{align}
  h_+(t)=\mathcal{A}(1+\cos^2\iota)\left[\cos\left(2\pi f t+\pi\dot f t^2+\phi\right)+\frac{9e}{4}\cos\left(2\pi f' t+\pi\dot f' t^2+\phi'\right)+\ldots\right]\text{,}
  \label{eq1}
\end{align}
where the ``ellipses'' refer to neglected terms at first and higher orders in eccentricity. $\mathcal{A}$ is the amplitude of the GWs signal (it varies as $1/a$) and $\iota$ is the inclination of the orbit with respect to the plane which is orthogonal to the direction of propagation of the signal propagating towards the observer. $\phi$ and $\phi'$ are the phases at the origin $(t=0)$ and are given by $\phi=2L_0$ and $\phi'=3L_0-\varpi_0$, where $L_0$ and $\varpi_0$ are the initial mean longitude and initial longitude of the pericenter, respectively. $f$ and $\dot f$ are called the main frequency and the main frequency shift, respectively. They are related to the dynamics of the binary systems according to the following expressions:
\begin{equation}
  2\pi f=2n\left(1+\frac{\dot L_{\mathrm{GR}}}{n}+\frac{\dot L_{\mathrm{M}}}{n}\right)\text{,} \qquad \pi\dot f=\frac{3n}{2}\,\frac{|\dot a_{\mathrm{GR}}|}{a}\text{,}
  \label{eq2}
\end{equation}
where the terms with the subscript ``GR'' refer to a secular contribution from general relativity (the expressions of these terms can be found in \citet{2022PhRvD.105l4042B}) and where $n$ is the mean motion of the orbit as given by Kepler third law of motion, namely $n=(Gm/a^3)^{1/2}$.

The terms $f'$ and $\dot f'$ in Eq. \eqref{eq1} are the frequency and the frequency shift of the first harmonic, respectively. As for the main frequency, they are related to the dynamics of the binary systems; they read as
\begin{equation}
  2\pi f'=3n\left(1+\frac{3\dot L_{\mathrm{GR}}-\dot\varpi_{\mathrm{GR}}}{3n}+\frac{3\dot L_{\mathrm{M}}-\dot\varpi_{\mathrm{M}}}{3n}\right)\text{,} \qquad \pi\dot f'=\frac{9n}{4}\,\frac{|\dot a_{\mathrm{GR}}|}{a}\text{.}
  \label{eq3}
\end{equation}

\begin{figure}
  \centering
  \includegraphics[width=0.46\textwidth,clip]{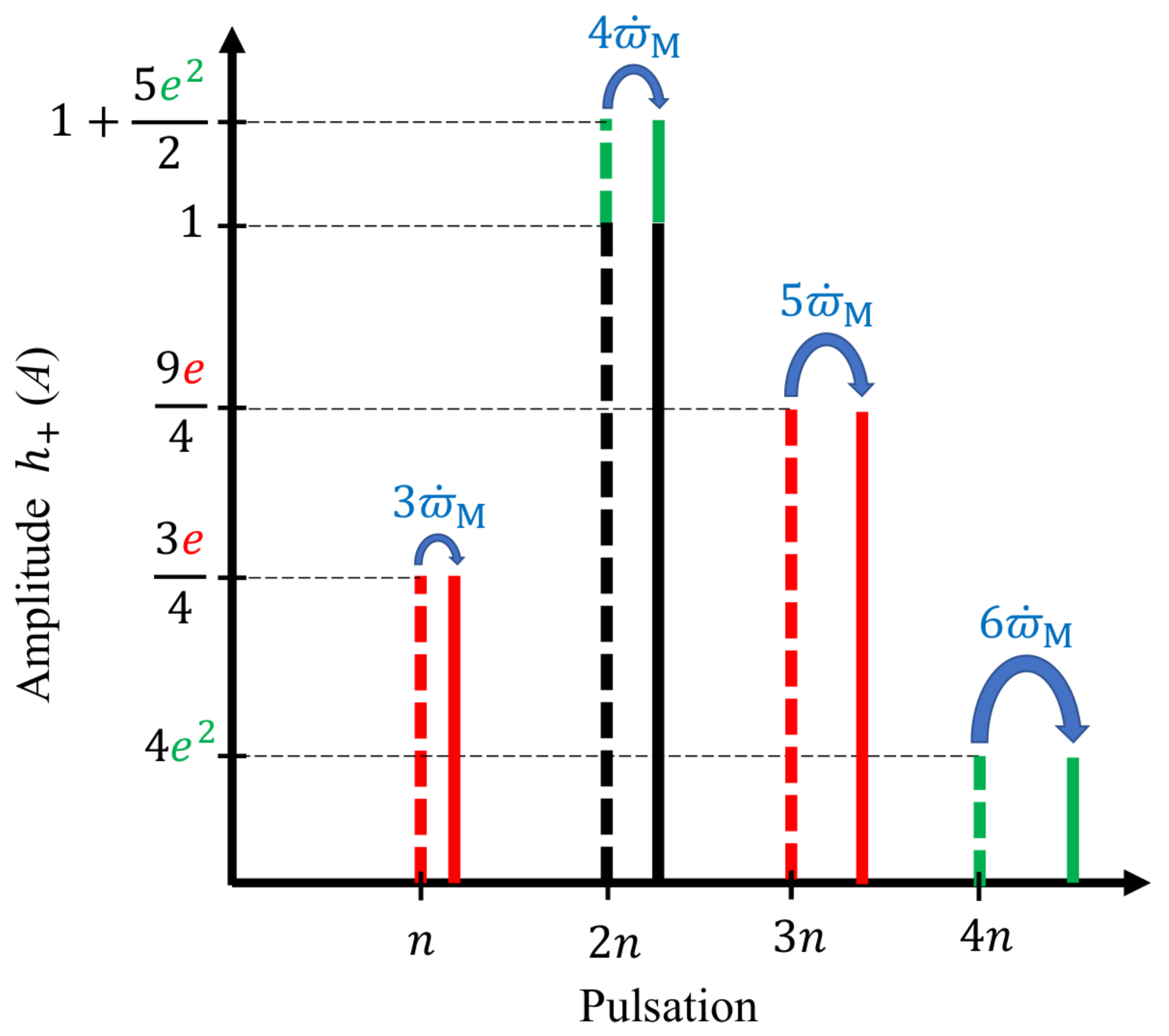}      
%% Note the ABSENCE of the extension .pdf  !
  \vspace{-0.1cm}
  \caption{Illustration of the frequency spectrum of the GWs mode polarization $h_+$ emitted by a magnetic binary system in quasi-circular orbit. The \emph{dashed} vertical lines represent the frequency peaks that would be observed for a binary system in quasi-circular orbit without magnetism. On the contrary, the \emph{thick plain} vertical lines represent the same spectrum considering the magnetic effect. The \emph{black}, \emph{red}, and \emph{green} colors represent the zeroth, first, and second order terms in eccentricity, respectively. The LDC quasi-monochromatic picture corresponds to the \emph{thick black} vertical line only.}
  \label{fig1}
\end{figure}

The frequency spectrum of the GWs mode polarization $h_+$ (see Eq. \ref{eq1}) is represented schematically in figure~\ref{fig1}. If we assume for a moment that the binary system is following a Keplerian orbit, the main frequency and the frequency of the first harmonic reduce respectively to $f=2P^{-1}$ and $f'=3P^{-1}$ with $P$ the period of the orbit. In these conditions, all effects other than the Newtonian acceleration between the two point-masses vanish so that $\dot f=0$ and $\dot f'=0$. This shows that a non-null eccentricity generates a GWs signal which is a superposition of monochromatic signals where the amplitude of the first harmonic is proportional to the eccentricity: $\tfrac{9e}{4}\mathcal{A}$ with $\mathcal{A}$ being the amplitude of the main frequency peak. If the binary system is not anymore in Keplerian orbit, the frequencies of the signal are shifted by both general relativity (1PN terms) and magnetic effects (see Eqs.~\eqref{eq2} and \eqref{eq3}). General relativity (2.5PN terms) also induces a drift of the frequencies due to the change in the semi-major axis caused by the emission of GWs. Because the rate of change of the semi-major axis is slow for inspiral GBs, the full GWs signal becomes a superposition of quasi-monochromatic signals. If we now focus on the effect of the magnetic dipole-dipole interaction, we infer from Eqs.~\eqref{eq2} and \eqref{eq3} that magnetism shifts the frequency (more precisely the pulsation) $kn$ with $k\in\mathbb{N}^+\setminus\{0\}$ by the amount: $(k+2)\dot\varpi_{\mathrm{M}}$ with respect to the case without magnetism. Hence, if LISA is able to measure the main frequency of the GWs signal with the precision $\sigma_f$ that is lower than the magnetic shift, that is to say
\begin{align}
  \left(\frac{\sigma_f}{f}\right)&<6.8\times10^{-7}\left(\frac{f}{10^{-1}\,\mathrm{Hz}}\right)^{4/3}\left(\frac{1.2\,\mathrm{M}_\odot}{m_1}\right)\left(\frac{0.3\,\mathrm{M}_\odot}{m_2}\right)\nonumber\\
  &\times\left(\frac{B_1}{10^9\,\mathrm{G}}\right)\left(\frac{B_2}{10^9\,\mathrm{G}}\right)\left(\frac{R_1}{6\times10^3\,\mathrm{km}}\right)^3\left(\frac{R_2}{15\times10^3\,\mathrm{km}}\right)^3\text{,}
\end{align}
then, one should be careful while attempting to interpret the measured frequency directly in term of the binary's masses since part of the frequency might have a magnetic origin. In the last inequality, we assumed a binary made of two highly magnetic white dwarfs, assuming Eq. (3) of \citet{2019MNRAS.488...64P} as expressions for both magnetic moments with $B$ and $R$ denoting the magnetic field and radius, respectively. Let us emphasize that verification binaries are actually known with relative uncertainties ranging from $10^{-6}$ to $10^{-9}$ \citep{2018A&A...616A...1G}, and we expect LISA to reach even higher precision.

\section{Data processing using the LDC quasi-monochromatic picture}

Currently the data processing pipeline developed in the LDC considers that the GWs signal from GBs is quasi-monochromatic (cf. Eq. \eqref{eq1} with $e=0$). We now want to explore how the total GWs signal in Eq. \eqref{eq1} (i.e., with $e\neq 0$), which corresponds to the case of a magnetic GB in quasi-circular orbit, would be processed by the quasi-monochromatic LDC tool. To do this, we simulate 4 years of LISA data by generating the three time delay interferometry channels (i.e., $A$, $E$, and $T$) from the GWs mode polarizations $h_+$ and $h_\times$ derived in the previous section (cf. Eq. \eqref{eq1} with $e=0.1$ while considering all terms up to $\mathcal{O}(e^2)$). Then, the simulated data are analyzed using LDC ``fastGB'' algorithm for quasi-monochromatic GB. 

\begin{figure}
  \centering
  \includegraphics[width=0.496\textwidth,clip]{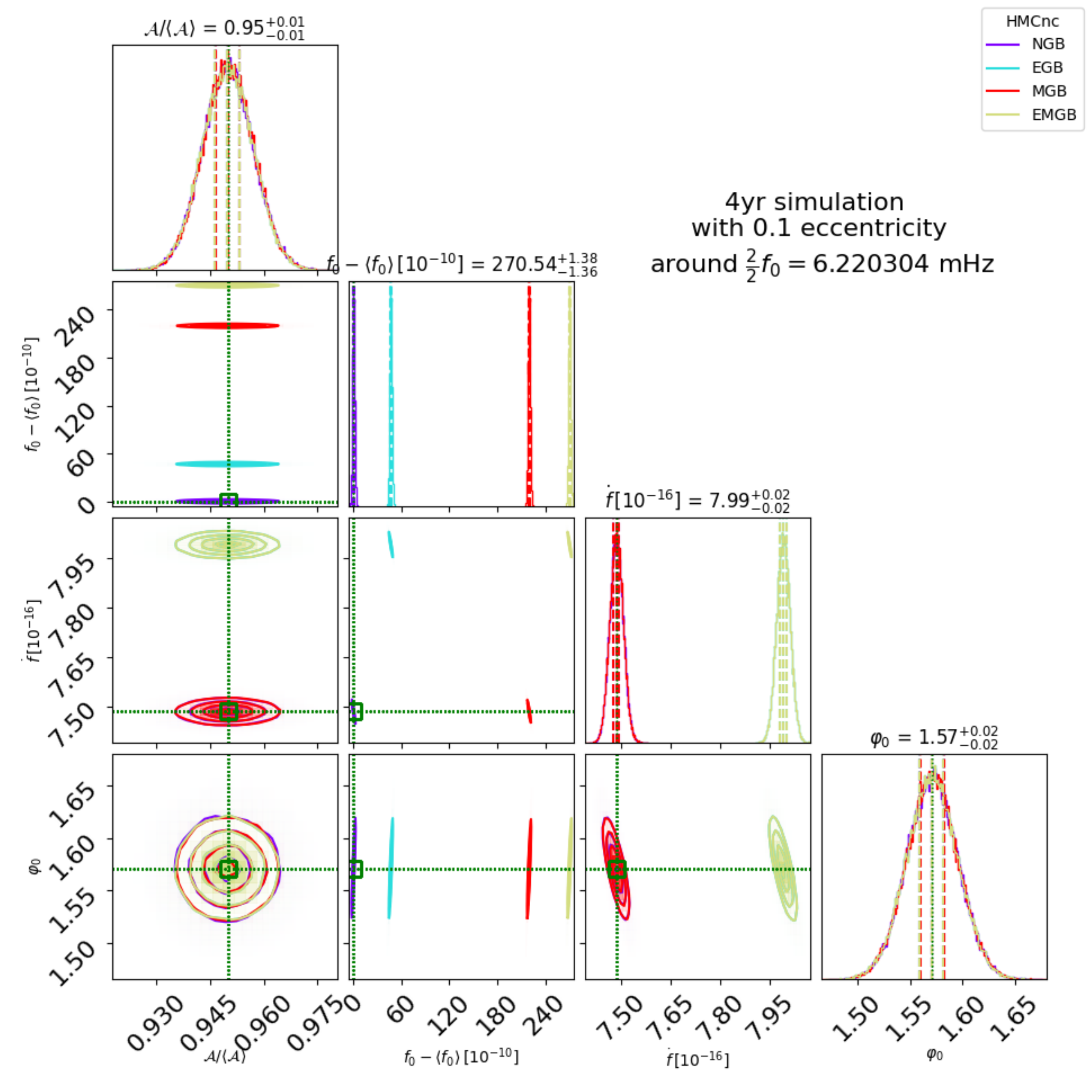}
  \includegraphics[width=0.496\textwidth,clip]{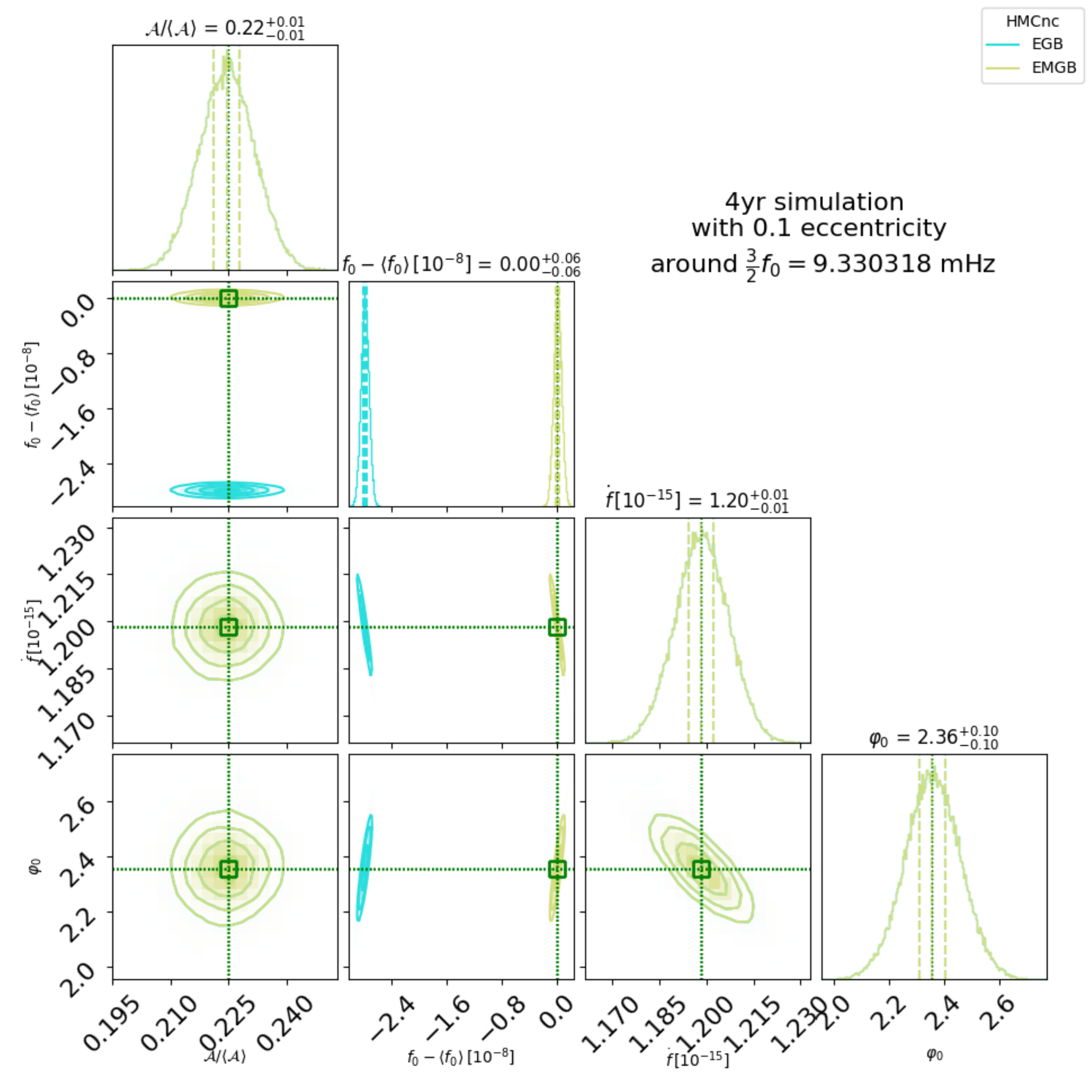}
  \vspace{-0.1cm}
  \caption{Parameter estimates from LDC quasi-monochromatic algorithm of 4 years simulated LISA data. The source of the GWs is a magnetic GB in quasi-circular orbit with $e=0.1$. \emph{Left}: parameter estimates of the main frequency peak (at $2n$). \emph{Right}: parameter estimates of the main harmonic (at $3n$). ``NGB'', ``EGB'', ``MGB'', and ``EMGB'' stand for ``non-magnetic GB'', ``eccentric GB'', ``magnetic GB'', and ``eccentric magnetic GB'', respectively.}
  \label{fig2}
\end{figure}

As can be seen in the left panel in figure \ref{fig2}, the central peak of the gravitational wave is easily identified and all parameters can be correctly estimated. The good uncertainties obtained on the central frequency and the frequency shift allow us to disentangle between the magnetic and the eccentric effects. Thus, a lack of knowledge on the eccentricity or magnetism could lead to a bias in the catalog of GBs. The frequency of the first harmonic of the GWs can also be detected by the algorithms (see the right panel in figure \ref{fig2}) and provides a suitable parameter estimate. With current algorithms, this peak could be interpreted as a new source with a lower amplitude and frequency. Again, this could biased the GBs' catalog.

\section{Conclusion}
%%--------------------

In this work, we first derived the secular orbital motion of a GB in quasi-circular orbit considering general relativity up to the 2.5PN approximation and the magnetic dipole-dipole interaction. Then, we derived the expressions of the mode polarizations. The total GWs signal is thus a superposition of quasi-monochromatic signals. From this signal, we simulated 4 years of LISA data and analyzed it with LDC tools. We showed that the first harmonic at $3n$ is detected as a new GB which can biased population studies from future LISA catalog. 

% Optional acknowledgements
% -------------------------
\begin{acknowledgements}
A.B. is grateful to Centre National d'\'Etudes Spatiales (CNES) for financial support. This work was supported by the Programme National GRAM of CNRS/INSU with INP and IN2P3 co-funded by CNES.
\end{acknowledgements}

\bibliographystyle{aa}  % A&A bibliography style file (aa.bst)
\bibliography{Bourgoin_S13} % your references in file: Yourfile.bib

\begin{thebibliography}{11}
\expandafter\ifx\csname natexlab\endcsname\relax\def\natexlab#1{#1}\fi

\bibitem[{{Amaro-Seoane} {et~al.}(2017){Amaro-Seoane}, {Audley}, {Babak},
  {Baker}, {Barausse}, {Bender}, {Berti}, {Binetruy}, {Born}, {Bortoluzzi}, \&
  {et. al}}]{2017arXiv170200786A}
{Amaro-Seoane}, P., {Audley}, H., {Babak}, S., {et~al.} 2017

\bibitem[{{Babak} {et~al.}(2020){Babak}, {Le Jeune}, {Petiteau}, \&
  {Vallisneri}}]{LDCGroup}
{Babak}, S., {Le Jeune}, M., {Petiteau}, A., \& {Vallisneri}, M. 2020,
  LISA-LCST-SGS-MAN-001, Rev. 1

\bibitem[{{Bagnulo} \& {Landstreet}(2022)}]{2022ApJ...935L..12B}
{Bagnulo}, S. \& {Landstreet}, J.~D. 2022, \apjl, 935, L12

\bibitem[{{Blanchet}(2014)}]{2014LRR....17....2B}
{Blanchet}, L. 2014, Living Reviews in Relativity, 17

\bibitem[{{Bourgoin} {et~al.}(2022){Bourgoin}, {Le Poncin-Lafitte}, {Mathis},
  \& {Angonin}}]{2022PhRvD.105l4042B}
{Bourgoin}, A., {Le Poncin-Lafitte}, C., {Mathis}, S., \& {Angonin}, M.~C.
  2022, \prd, 105, 124042

\bibitem[{{Ferrario} {et~al.}(2020){Ferrario}, {Wickramasinghe}, \&
  {Kawka}}]{2020AdSpR..66.1025F}
{Ferrario}, L., {Wickramasinghe}, D., \& {Kawka}, A. 2020, Advances in Space
  Research, 66, 1025

\bibitem[{{Ferrario} \& {Wickramasinghe}(2005)}]{2005MNRAS.356..615F}
{Ferrario}, L. \& {Wickramasinghe}, D.~T. 2005, \mnras, 356, 615

\bibitem[{{Gaia Collaboration} {et~al.}(2018){Gaia Collaboration}, {Brown},
  {Vallenari}, {Prusti}, {de Bruijne}, {Babusiaux}, {Bailer-Jones}, {Biermann},
  {Evans}, {Eyer}, \& et~al.}]{2018A&A...616A...1G}
{Gaia Collaboration}, {Brown}, A.~G.~A., {Vallenari}, A., {et~al.} 2018, \aap,
  616, A1

\bibitem[{{Pablo} {et~al.}(2019){Pablo}, {Shultz}, {Fuller}, {Wade}, {Paunzen},
  {Mathis}, {Le Bouquin}, {Pigulski}, {Handler}, {Alecian}, {Kuschnig},
  {Moffat}, {Neiner}, {Popowicz}, {Rucinski}, {Smolec}, {Weiss}, {Zwintz}, \&
  {BinaMIcS Collaboration}}]{2019MNRAS.488...64P}
{Pablo}, H., {Shultz}, M., {Fuller}, J., {et~al.} 2019, \mnras, 488, 64

\bibitem[{{Shultz} {et~al.}(2015){Shultz}, {Wade}, {Alecian}, \& {BinaMIcS
  Collaboration}}]{2015MNRAS.454L...1S}
{Shultz}, M., {Wade}, G.~A., {Alecian}, E., \& {BinaMIcS Collaboration}. 2015,
  \mnras, 454, L1

\bibitem[{Timpano {et~al.}(2006)Timpano, Rubbo, \&
  Cornish}]{PhysRevD.73.122001}
Timpano, S.~E., Rubbo, L.~J., \& Cornish, N.~J. 2006, \prd, 73, 122001

\end{thebibliography}

\end{document}